                     \def\t{\tau}

\documentstyle[abstract,aps,eqsecnum]{revtex}

\def\be{\begin{equation}}             \def\ee{\end{equation}}
\def\bea{\begin{eqnarray} }           \def\eea{\end{eqnarray} }
\def\ba#1{\begin{array}{#1}}          \def\ea{\end{array}}
                
\def\bsubeq{\begin{mathletters}}      \def\esubeq{\end{mathletters}}
\def\bitem{\begin{itemize}}           \def\eitem{\end{itemize}}

\begin{document}
\draft

\title{Canonical approach to 2D induced gravity} 

\author{D. S. Popovi\'c\thanks{E-mail address: popovic@phy.bg.ac.yu} and B. Sazdovi\'c\thanks{E-mail address: sazdovic@phy.bg.ac.yu}}
\address{Institute of Physics, 11001 Belgrade, P.O.Box 57, Yugoslavia}
\date{\today}
\maketitle

\begin{abstract}

Using canonical method the Liouville theory has been obtained as
a gravitational Wess-Zumino action of the Polyakov string. From
this approach it is clear that the form of the Liouville action is
the consequence of the bosonic representation of the Virasoro
algebra, and that the coefficient in front of the action is proportional
to the central charge and measures the quantum braking of the
classical symmetrie.

\end{abstract}

\pacs{PACS number(s): 11.10.Kk, 11.10.Ef}

\section{Introduction}

In the paper [1] we introduced new canonical method to investigate
anomalous gauge theories. By definition these are the theories
where symmetries of the classical action are broken by the quantum
effects. In the hamiltonian language it means that the classical
theory has first class constraints (FCC) $t_m$, satisfying closed
Poisson bracket (PB) algebra, and that after quantization the
corresponding commutator algebra of the operators $\hat t_m$
obtains the central charge and consequently constraints become the
second class (SCC). The basic idea of the new approach is to
introduce the variables $T_m$ and postulate that their classical
PB algebra is isomorphic to the quantum commutator algebra of $\hat
t_m$. This defines the effective theory. To find the effective
action of this theory we first parameterize the constraints $T_m$
by some phase-space variables and "solve" the PB constraints
obtaining the expressions for $T_m$ in terms of coordinates and
momenta. Then we use the general canonical formalism [2,3,1]
connecting the action $I$ with the expressions for the
hamiltonian $H_0$ and the primary constraints $T_m$ in terms of
the phase-space coordinates. It states that the action
\be
I=\int d^2\xi (p_i{\dot q}^i-H_0-h^mT_m) \, ,
\ee
is invariant under gauge transformations
\be
\delta R=\varepsilon^m\{ R, T_m\} ,\qquad \delta
h^m={\dot\varepsilon}^m-h^n\varepsilon^kU_{kn}^{\,\,\,\,
m}-\varepsilon^nV_n^{\,\,\, m}  \, ,
\ee
of any quantity $R(p, q)$ and lagrange  multipliers
$h^m$ if the constraints $T_m$ are FCC satisfying PB algebra

\be \{ T_m, T_n\} =U_{mn}{}^{k}T_k,\qquad \{ H_0, T_m\}
=V_m^{\,\, n}T_n \,  .  \ee
If $T_m$ are SCC, then instead of the first eq. (1.3) we
have
\be
\{ T_m, T_n\} =U_{mn}{}^{k} T_k+\Delta_{mn} \, ,
\ee
and the variation of the action is proportional to the
Schwinger term $\Delta_{mn}$
\be
\delta I=-\int d\t h^m\Delta_{mn}\varepsilon^n \, .
\ee
We will use this statement in both directions, to find
the symmetry transformations from the known action, and the more
important to us, to construct the effective action from the
constraints in terms of the coordinates and momenta.

In the paper [1] we started with the non-Abelian symmetry of the
classical action whose generators satisfy the Kac-Moody algebra. The symmetry
has an axial anomaly and we obtained the gauged WZNW model as an
effective theory.

In this paper we are going to replace the non-Abelian symmetry with
the diffeomorphysms. We expect to obtain Virasoro algebra as
an algebra of constraints, and reparameterization or trace anomalies,
depending on the regularization procedure. As an effective action
we will get a Liouville action. We choose the Polyakov string
theory [4] as an example. The quantization of this theory has been
obtained by path integral approach, which in the case of string
becomes a sum over random surfaces [4,5]. The group of authors [6]
also develops a canonical treatment of 2D gravity using
Batalin, Fradkin and Vilkovisky method. We believe that our
approach is not just one more in the series but the simplest one,
and besides works naturally  in the hamiltonian formalism.

In Sec. II we introduce the Polyakov string. We fix the
reparameterization symmetry and make a canonical analysis. All
constraints are of the FCC and as a consequence of the
reparameterization symmetry the generators of the scalar and
Faddeev-Popov (FP) parts satisfy the same Virasoro algebra, but they
have different representations.

In Sec. III we quantize  the theory, introducing operators instead
of the fields, and presence of the central terms changes the nature of
the constraints  from the FCC to the SCC. The only difference
between scalar and FP part appears in the values of the central
charges $c^x$ and $c^{gh}$, which is the only remnant of the original
system.

Sec. IV is a central part of the paper. We find an expression for the
constraints, in terms of phase-space coordinates, from the
requirement that its PB algebra is isomorphic to the commutator
algebra of the Sec. III.  With the help of such bosonic
representation of the Virasoro algebra we construct the effective
action. Elimination of the momentum variables on their equations
of motion yields the Liouville action. This is the gravitational
Wess-Zumino (WZ) action because it measures the breaking of the
reparameterization invariance. By adding some finite local
counterterm we obtain the action which is invariant under
diffeomorphisms but not under Weyl rescaling or in the other words
we shift the anomaly from the Virasoro to the trace one.

Sec. V is devoted to concluding remarks.

\section{Canonical analysis of the Polyakov string}

\subsection{Constraints in the scalar theory}

Let us consider the action for Polyakov string $S_0$

\be
S_0(x^M, g_{\mu\nu})={1\over 2}\int
d^2\xi\sqrt{-g}g^{\mu\nu}\partial_\mu x^M\partial_\nu x_M \, ,
\ee
where $M=0, 1, ..., D-1$, and $\mu , \nu =0, 1$.
One can rewrite the action as
\be
S_0(x^M, h)=\int d^2\xi {1\over
h^--h^+}(\partial_0+h^-\partial_1)x^M(\partial_0+h^+\partial_1)x^M \, ,
\ee
in terms of the light-cone variables $h^-, h^+$ and $F$,
defined by the expression
\be
g_{\mu\nu}=e^{2F}{\hat g}_{\mu\nu}={1\over
2}e^{2F}\left|\begin{array}{cc}-2h^-h^+& \qquad h^-+h^+\\
h^-+h^+&\qquad -2\end{array} \right| \, .
\ee
As a consequence of the conformal symmetry the action (2.2) does
not depend on $F$.

The canonical momenta are
\be
p_M={\delta S_0\over \delta {\dot x}^M}={1\over
h^--h^+}[2{\dot x}^M+(h^-+h^+)x^{M\prime}]\, ,
\ee
\be
p_\pm ={\delta S_0\over\delta {\dot h}^\pm}=0 \,  ,
\ee
where ${\dot X}=\partial_0 X$ and $X^{\prime} =\partial_1 X$ for any variable $X$.
The canonical hamiltonian density ${\cal H}_c$ can be
expressed in terms of the currents
\be
j_\pm^M={p^M\pm x^{M\prime}\over\sqrt{2}} \, ,
\ee
 as
\be
{\cal H}_c=h^+t_+^x+h^-t_-^x ,\qquad t_\pm^x=\mp{1\over
2}j_\pm^M j_{\pm M} \, .
\ee
The index $x$ denotes that this energy-momentum tensor
belongs to the matter fields $x^M$.

Starting with the basic PB $\quad  \{ x^M(\sigma ), p_N
({\bar\sigma})\} =\delta_{\, \, N}^M\delta (\sigma -{\bar\sigma})\quad$
we have
\be
\{ j_\pm^M, j_{\pm N}\} =\pm\delta_{\, \, N}^M\delta^\prime
,\qquad \{ j_\pm^M, j_{\mp N}\} =0 \, ,
\ee
which implies
\be
\{ t_\pm^x, j_\pm^M\}=-j_\pm^M \delta^\prime ,\qquad \{
t_\pm^x , j_{\mp}^M\}=0  \, ,
\ee
and
\be
\{ t_\pm^x (\sigma ), t_\pm^x({\bar\sigma})\} =-[t_\pm^x
(\sigma )+t_\pm^x ({\bar\sigma})]\delta^\prime (\sigma
-{\bar\sigma}),\qquad \{t_\pm^x, t_\mp^x\}=0 \, .
\ee
In relations (2.8-2.10) we recognize the semidirect
product of $D$ independent abelian Kac-Moody (KM) algebras (2.8)
and Virasoro algebras (2.10).

We introduce the total hamiltonian
\be
H_T=\int d\sigma [{\cal H}_c+\lambda^+p_++\lambda^-p_-] \, ,
\ee
and from the consistency condition for the primary
constraints $p_\pm$
\be
{\dot p}_\pm =\{ p_\pm , H_T\}=-t_\pm^x \,
\ee
we conclude that $t_\pm^x$ are secondary constraints.

With the help of PB (2.10) it is clear that all constraints
$p_\pm$ and $t_\pm^x$ are FCC, and that there are
no more constraints. The canonical hamiltonian ${\cal H}_c$ is
weakly equal to zero.

We can write action (2.2) in the hamiltonian form
\be
S_0(x^M, p_M , h)=\int d^2\xi (p_M{\dot
x}^M-h^+t_+^x-h^-t_-^x) \, ,
\ee
which is convenient for obtaining the local symmetries. It
is easy to check that on the equation of motion for the momenta
$p_M$ the action (2.13) yields (2.2).

Comparing (2.13) with (1.1) and (2.10) with first equation (1.3) we can conclude
from second equation (1.2) that
\be
\delta h^\pm
=(\partial_0+h^\pm\partial_1-h^{\pm\prime})\varepsilon_\mp
=\sqrt{2}\sqrt{{-\hat g}}\mathop{\hat\nabla}\limits_{\mp
1}{}_\mp\varepsilon_\mp \, ,
\ee
where the covariant derivative on tensor $V_n$
($n\in Z$ stands for the sum of the indices taking 1 for plus and -1 for
minus) is
$ \mathop{\hat\nabla}\limits_{n}{}_\pm V_n = ({\hat\partial}_\pm
+n{\hat\omega}_\pm) V_n \,$
with  ${\hat\partial}_\pm={\sqrt{2}\over
h^--h^+}(\partial_0+h^\mp\partial_1)$ and ${\hat\omega}_\pm
={\mp\sqrt{2}\over h^--h^+}h^{\mp\prime}$.

With the help of (2.3) we recognize (2.14) as the diffeomorphism
transformation of the metric density ${\tilde
g}^{\mu\nu}=\sqrt{-g}g^{\mu\nu}$, after introducing new parameters
$\varepsilon^\pm =\varepsilon^1 -h^\pm\varepsilon^0$, see [3] for
more details.

\subsection{Gauge fixing and constraints in Faddeev-Popov action}

It is well known that classically we can gauge away all components
of the metric tensor. Due to the presence of the anomaly at the quantum
level, we are not allowed to fix all these symmetries at the classical level.
Let us fix the diffeomorphism using BRST method.

To obtain the BRST transformations we replace gauge parameters
with ghost fields
\be
\varepsilon_\pm\to c_\pm \,  ,
\ee
so that instead of (2.14) we have
\be
s h^\pm =\sqrt{2}\sqrt{-{\hat g}} \mathop{\hat\nabla}\limits_{\mp 1}{}_\mp c_\mp \,  .
\ee
We introduce antighosts ${\bar c}_{\pm\pm}$  and
auxiliary fields $b_{\pm\pm}$ with the BRST transformations
\be
s {\bar c}_{\pm\pm}=b_{\pm\pm}, \qquad sb_{\pm\pm}=0 \, ,
\ee
and choose the gauge fermion $\Psi$ in the form
\be
\Psi ={\bar c}_{++}(h^+-h_b^+)+{\bar c}_{--}(h^--h_b^-) \,  ,
\ee
to fix the gauge fields $h^\pm$ to some background fields $h^\pm_{b}$.

Starting from the expression

\be
-s\Psi ={\cal L}_{GF}+{\cal L}_{FP} \, ,
\ee
one can integrate over auxiliary fields $b_{\pm\pm}$,
and then over $h^\pm$ obtaining $h^\pm=h_b^\pm$, and the FP action
\bea
S_{FP}=\int d^2\xi {\cal L}_{FP}&=& \sqrt{2}\int
d^2\xi\sqrt{-{\hat g}}({\bar
c}_{++}\mathop{\hat\nabla}\limits_{-1}{}_-c_-+{\bar
c}_{--}\mathop{\hat\nabla}\limits_1{}_+c_+)  \,\nonumber \\
&=& \int d^2\xi [{\bar
c}_{++}(\partial_0+h^+\partial_1-h^{+\prime})c_{-}+{\bar
c}_{--}(\partial_0+h^-\partial_1-h^{-\prime})c_+] \,  .
\eea

From now on only the background fields $h_b^\pm$ exist, and
for simplicity we will omit index $b$ and write simply $h^\pm$.

This action is already in the hamiltonian form. The coordinates
are ghosts $c_{\pm}$, the conjugate momenta
\be
\pi _{\pm \pm}={\delta_L S_{FP} \over\delta{\dot
c}_{\mp }}=-{\bar c}_{\pm \pm} \, ,
\ee
are antighosts and the energy-momentum tensor
\be
 t^{gh}_\pm =
{\delta S_{FP} \over \delta h^\pm}=
{\bar c}_{\pm \pm}c_{\mp }^\prime +({\bar c}_{\pm \pm}c_{\mp })^\prime =
2{\bar c}_{\pm \pm }c_{\mp }^\prime +{\bar c}_{\pm \pm}^\prime
c_{\mp n} \, ,
\ee
plays the role of the primary constraints corresponding
to the lagrange multipliers $h^\pm$. In terms of the momenta we
obtain
\be
 t^{gh}_\pm =
-2\pi_{\pm \pm}c_{\mp }^\prime - \pi^\prime_{\pm \pm}c_{\mp }  \, .
\ee

Starting with the basic PB for $c$ and $\pi$
\be
\{ c_\mp (\sigma ), \pi_{\pm \pm}({\bar\sigma})\} =\delta (\sigma
-{\bar\sigma}) \,  ,
\ee
one can find that PB of $t^\prime$s satisfies two
independent copies of Virasoro algebras without central charges,
\be
\{ t_\pm^{gh} (\sigma ), t_\pm^{gh} ({\bar\sigma})\}
=-[t_\pm^{gh} (\sigma )+t_\pm^{gh}
({\bar\sigma})]\delta^\prime (\sigma -{\bar\sigma}), \qquad \{
t_+^{gh} , t_-^{gh} \} =0 \, .
\ee

Note that the Virasoro algebras for matter fields (2.10) and for
the ghosts (2.25) are classically identical. The different
expressions for the energy momentum tensors $t^x_\pm$ (2.7)
and $t^{gh}_\pm$ (2.23) will cause the different quantum algebras.

\section{Canonical quantization of the Polyakov string}

In this section we are going to perform quantization of the matter
fields $(x^M)$ and the ghost fields $(c_\pm ,{\bar c}_{\pm\pm})$ in the
gauge fixed Polyakov string theory, with the action
\be
S=S_0(x, h)+S_{FP}({\bar c}, c, h) \, .
\ee
The method developed in [1] for non-Abelian gauge symmetry will be
applied here for reparameterization symmetry.

Transition from the classical to the quantum theory is achieved by
introducing the operators $\mathop{\hat\Omega}_\pm =\{ {\hat
j}_\pm , {\hat c}_\mp , {\hat \pi}_{\pm \pm}\}$ instead of the fields
$\Omega_\pm =\{ j_\pm , c_\mp , \pi_{\pm \pm} \}$, replacing the PB (2.8)
and (2.24) by the (anti)commutators
\be
[{\hat j}_\pm^M, {\hat j}_{\pm N}]=\pm
i\hbar\delta_N^M\delta^\prime ,\qquad [{\hat c}_\mp(\sigma
),{\hat\pi}_{\pm \pm}(\bar\sigma)]=i\hbar\delta \, ,
\ee
and defining the composite operators using normal
ordering prescription
\be
{\hat t}_\pm^x=\mp {1\over 2}:{\hat j}_\pm^M{\hat j}_{\pm M}:\,\,\, ,\qquad
{\hat t}_\pm^{gh} =
-2 :{\hat \pi}_{\pm \pm}{\hat c}_\mp^\prime :-:{\hat\pi}_{\pm \pm}^\prime
{\hat c}_\mp : \,\, .
\ee

In order to obtain commutator algebra for $t^\prime$s we decompose
operators in positive and negative frequencies in the position
space [7,1]
\be {\hat \Omega}^{(\pm )}(\tau , \sigma
)=\int\limits_{-\infty}^{+\infty}d{\bar\sigma}\delta^{(\pm
)}(\sigma -{\bar\sigma}){\hat \Omega}(\tau ,{\bar\sigma}), \ee
where delta function is given by
\bea
\delta (\sigma )&=&\delta^{(+)}(\sigma )+\delta^{(-)}(\sigma
)\, ,\\
 \delta^{(\pm)} (\sigma )&=&\int\limits_{-\infty}^{+\infty}{dk\over
2\pi}\theta (\mp k)e^{ik(\sigma\mp i\varepsilon )}={\pm i\over
2\pi (\sigma\pm i\varepsilon )}\,  , \qquad (\varepsilon
>0) \,  .
\eea
We define ${\hat \Omega}_\pm^{(\mp )}$ as creation
operators and ${\hat\Omega}_\pm^{(\pm )}$ as annihilation
operators
\be
<0|{\hat\Omega}_\pm^{(\mp )}=0 \, , \qquad {\hat\Omega}_\pm^{(\pm)} |0>=0 \, .
\ee

With this choice we preserve the symmetry under parity
transformations on the quantum level. It corresponds to the
left-right symmetric regularization scheme and consequently we
will obtain anomaly for both Virasoro algebras. Converting the
Virasoro anomaly to the trace anomaly is possible due to finite local
counterterm (Sec. IV C).

Up to central terms $\Delta_\pm$, the commutator algebra for both
matter and ghost fields should be the same,
\bea
&[{\hat t}_\pm& , {\hat t}_\pm ]=-i\hbar [{\hat t}_\pm
(\sigma )+{\hat t}_\pm ({\bar\sigma})]\delta^\prime (\sigma -{\bar
\sigma} )+\Delta_\pm (\sigma ,{\bar\sigma}), \\ &[{\hat t}_+&,
{\hat t}_-]=0 \, .
\eea
To find the central term we take the vacuum expectation value of
$[{\hat t}_\pm , {\hat t}_\pm ]$. Note that under parity
transformation $P: {\hat\Omega}_+(\tau ,\sigma )\to
{\hat\Omega}_-(\tau ,-\sigma)$ so that $P {\hat t}_\pm(\tau
,\sigma )P=-{\hat t}_\mp (\tau ,-\sigma )$. Consequently we have
$\Delta_+=-\Delta_-$ and it is enough to calculate only
$\Delta_+$. From the basic commutation relations (3.2) we find
\be
[{\hat j}_{+}^{(\pm )M},{\hat j}_{+N}^{(\mp )}]=
i\hbar\delta_N^M\delta^{(\pm )^\prime} \, ,
\ee
and
\be
[{\hat c}_-^{(\pm )}, {\hat\pi}_{++}^{(\mp )}]=i\hbar\delta^{(\pm)}\, ,\qquad
[{\hat c}_-^{(\pm )^\prime}, {\hat\pi}_{++}^{(\mp)}]=i\hbar\delta^{(\pm)^\prime}\, , \qquad
[{\hat c}_-^{(\pm )},{\hat\pi}_{++}^{(\mp )^\prime}]=-i\hbar\delta^{(\pm )^\prime} \, .
\ee
After straightforward calculation we obtain
\be
\Delta_\pm (\sigma , {\bar\sigma})=\pm i\hbar {\hbar\over
24\pi}c\delta^{\prime\prime\prime}(\sigma -{\bar\sigma}) \, ,
\ee
where the central charges for the matter and ghost
fields are respectively
\be
c^x=D,\qquad c^{gh} =-26 \,  .
\ee

\section{2D induced gravity as an effective action}

We want to find effective theory for the matter and ghost part of
the Polyakov action. Instead of the FCC PB algebras (2.10) and
(2.25) of diffeomorphism generators, we obtained the SCC
commutator algebras (3.8, 3.9). These algebras have the central
terms and therefore break the reparameterization symmetry and
become the algebras of dynamical variables.

\subsection{Bosonic representation of the Virasoro algebra}

The algebras for matter and ghost fields have the same structure up to
the central charges, and so we will investigate both cases together.
Following [1] we introduce new variables $T_\pm$ instead of ${\hat
t}_\pm$ and postulate their classical PB
\bea
&\{ T_\pm &, T_\pm\} =-[T_\pm (\sigma )+T_\pm
({\bar\sigma})]\delta^\prime
\pm\kappa_0c\delta^{\prime\prime\prime}, \qquad
(\kappa_0\equiv{\hbar\over 24\pi}) \\ &\{ T_+&, T_-\}=0 \, ,
\eea
to be isomorphic to the commutator algebras (3.8, 3.9)
of the operators ${\hat t}_\pm$. To find the solution of (4.1,
4.2) for $T_\pm$ as a function of the canonical bosonic variables $\varphi$ and
$\pi$ with the PB
\be
\{\varphi , \pi\} =\delta \, ,
\ee
we use the ansatz
\be
T_\pm =a_\pm K_\pm^2+b_\pm K_\pm^\prime +V_\pm \, ,
\ee
where $K_\pm =(\pi +\alpha_\pm\varphi^\prime )$ are
currents, $V_\pm =V_\pm (\varphi ,\pi )$  is momentum
dependent potential and $a_\pm , b_\pm$ and $\alpha_\pm$ are
constants.

The currents of opposite chirality should commute
\be
\{ K_+, K_-\} =0 \, ,
\ee
so we get $\alpha_+=-\alpha_- \equiv \alpha$ and
consequently
\be
K_\pm =(\pi\pm\alpha\varphi^\prime ) \, .
\ee
The currents of the same chirality satisfy abelian KM algebras
\be
\{ K_\pm , K_\pm\} =\pm 2\alpha\delta^\prime \, .
\ee

After some calculation we find that
\be
\{ T_\pm , T_\pm\}=\pm 4\alpha a_\pm \left\{ [T_\pm (\sigma )+T_\pm
({\bar\sigma })]\delta^\prime -{b_\pm^2\over
2a_\pm}\delta^{\prime\prime\prime} \right\} \, ,
\ee
if the potential is the solution of the equation
\be
V_\pm ={1\over 2}K_\pm\partial_\pi V_\pm \mp
{1\over2q_\pm}[\partial_\varphi V_\pm \mp\alpha (\partial_\pi
V_\pm )^\prime ] \, ,
\ee
 with
\be
q_\pm\equiv {2\alpha a_\pm\over b_\pm} \, .
\ee
Comparing (4.8) with (4.1) we conclude that the conditions
\be
\pm 4\alpha a_\pm =-1,\qquad 2\alpha b_\pm^2=-\kappa_0c \, ,
\ee
should be satisfied.

Let us solve the condition (4.9). Instead of $(\partial_\pi V_\pm
)^\prime$ we can write $\partial_\pi^2V_\pm\pi^\prime
+\partial_\varphi\partial_\pi V_\pm\varphi^\prime$ and because
$V_\pm$ does not depend on  $\pi^\prime$
and $\varphi^\prime$, respective coefficients should be equal to
zero. The first one
\be
\partial_\pi^2V_\pm =0 \, ,
\ee
implies that $V_\pm$ is linear in the momenta
\be
V_\pm =\pi u_\pm (\varphi )+\upsilon_\pm (\varphi ) \,  .
\ee
The second condition
\be
\partial_\pi (V_\pm\pm {1\over q_\pm}\partial_\varphi V_\pm
)=0 \,  ,
\ee
and the form of $V_\pm$ we have just obtained in(4.13)
give us the equation for $u$
\be
u_\pm\pm {1\over q_\pm}\partial_\varphi u_\pm =0 \, .
\ee
From remaining part of $V_\pm$ in (4.9) we obtain again (4.15) and
the  new equation
\be
v_\pm = \mp {1\over 2q}\partial_\varphi v_\pm \, .
\ee

Now $u_\pm$ and $v_\pm$ are given as
\be
u_\pm =u_{0\pm}e^{\mp q_\pm \varphi}, \qquad v_\pm
=v_{0\pm}e^{\mp 2q_\pm \varphi} \, ,
\ee
where $u_{0\pm}$ and $v_{0\pm}$ are constants and the
solution for $V_\pm$ is
\be
V_\pm =u_{0\pm}e^{\mp q_\pm \varphi}\pi +v_{0\pm}e^{\mp
2q_\pm \varphi} \, .
\ee

We can proceed in finding consequences of the relation $\{ T_+,
T_-\} =0$. All conditions connecting expressions of $V_+$ with
$V_-$  should be valid for any $\varphi$. So, we have
$\quad -q_+=q_-\equiv q$ and with the help of (4.10)
\be
{a_+\over b_+}=-{a_-\over b_-} \, .
\ee

After some calculation three groups of conditions are obtained,
because coefficients in front of $\delta^{\prime\prime},
\delta^\prime$ and $\delta$ must be zero. Any of these groups gives
few expressions because coefficients in front of $\pi , \pi^2,
\pi^\prime , \varphi^\prime , \varphi^{\prime^2},
\varphi^{\prime\prime}, \pi\varphi^\prime$ vanish separately. But
some conditions are giving relations already obtained, so we get only
four new relations
\bea
 a_+u_{0-} &-& a_-u_{0+}=0, \\ a_+v_{0-} &-& a_-v_{0+}=0,
\\ b_+u_{0-}&+&b_-u_{0+}=0,\\ b_+v_{0-} &+& b_-v_{0+}=0 \, .
\eea

Now from (4.11), (4.19) and (4.10) it is easy to conclude that
$a_+=-a_-={-1\over 4\alpha},\quad b_+=b_-\equiv b,\quad q={1\over
2b}, \quad u_{0+}=-u_{0-}\equiv {\theta\over 2q} ,\quad
v_{0+}=-v_{0-}\equiv\mu \quad$ and
\be
{\alpha\over 2q^2}=-\kappa_0c \, .
\ee

The expressions for $V_\pm$ and $T_\pm$ obtain the following form
\be
V_\pm =\pm ({\theta\over 2q} e^{q\varphi}\pi +\mu
e^{2q\varphi}) \, ,
\ee
and
\be
T_\pm ={\pm 1\over 8q^2\kappa_0c}K_\pm^2+{1\over
2q}K_\pm^\prime\pm\mu e^{2q\varphi}\pm{\theta\over 2q}
e^{q\varphi}\pi \, ,
\ee
where
\be
K_\pm =\pi\mp 2q^2\kappa_0c\varphi^\prime \, .
\ee

We can eliminate the parameter $q$ rescaling the canonical variables $\quad 2q\varphi
=\phi ,\quad  {1\over 2q}\pi =p$, and the currents $K_\pm =2qJ_\pm$ so
that
\be
\{ \phi ,p\} = \delta \, ,
\ee
and
\be
J_\pm =p\mp{\kappa_0c\over 2}\phi^\prime \, .
\ee

Finally we rewrite the solution of the equations (4.1) and (4.2)
in the form
\be
T_\pm ={\pm1\over 2\kappa_0c}J_\pm^2+J_\pm^\prime\pm \mu
e^\phi\pm\theta pe^{\phi /2} \, .
\ee

Let us stress that besides the well known terms $J^2, J^\prime$
and $e^\phi$, we have obtained the new one $\theta pe^{\phi /2}$,
linear in the momenta.

\subsection{Effective action}

Using the general canonical formalism described in the
introduction we are ready to derive the effective action
\be
W(\phi , p, h^+, h^-)=\int d^2\xi (p{\dot
\phi}-h^+T_+-h^-T_-) \, ,
\ee
knowing the expressions for the constraints $T_\pm$
(4.29-4.30).

To eliminate the momentum variable $p$, in order to obtain the
second-order form of the action, we consider the equation of
motion to be fulfilled
\be
{\dot\phi}-{1\over \kappa_0c}(h^+J_+-h^-J_-)+(h^++h^-)^\prime
+\theta (h^--h^+)e^{\phi /2}=0 \, .
\ee
With the help of (4.29) it yields
\be
J_\pm =-{\kappa_0c\over\sqrt{2}}[{\hat\partial}_\pm\phi
+{\hat\omega}_--{\hat\omega}_++\sqrt{2}\theta e^{\phi /2}] \, .
\ee
Substituting this back in (4.31) we obtain
\be
W(\phi , h^+, h^-)= -{\kappa_0 c\over 2} [W_L(\phi , h^+,
h^-)+W_\omega (h^+, h^-)+\theta W_\Omega (\phi ,h^+,h^-)] \,  ,
\ee
where
\be
W_L=\int d^2\xi\sqrt{-{\hat g}}
 \{ {\hat\partial}_+\phi{\hat\partial}_-\phi +{\hat R}
 \phi +M^2e^\phi \} ,\qquad \bigg( M^2=2\theta^2-{4\mu\over\kappa_0c}\bigg)
\ee
\be
W_\omega =\int d^2\xi\sqrt{-{\hat
g}}({\hat\omega}_--{\hat\omega}_+)^2=2\int d^2\xi\sqrt{-{\hat
g}}{\hat g}^{00}\bigg[\bigg( {{\hat g}_{01}\over {\hat
g}_{11}}\bigg)^\prime \bigg]^2 \, ,
\ee
and
\be
W_\Omega=2\sqrt{2}\int d^2\xi\sqrt{-{\hat g}}({\hat
 \partial}_+-{\hat\omega}_+ +
 {\hat \partial}_-+{\hat\omega}_-)e^{\phi/2}=\int
 d^2\xi\partial_\mu\Omega^\mu \, ,
\ee
with
\be
{\hat R} =2 {\hat\nabla}_- {\hat\omega}_+
-2 {\hat\nabla}_+ {\hat\omega}_- \, , \qquad
\Omega^0=4e^{\phi /2}, \qquad \Omega^1 =2(h^-+h^+)e^{\phi /2} \,  .
\ee

 The first term, $W_L$, is well known Liouville action. The
 cosmological term has two independent contributions, usual one
 proportional to $\mu$ and a new one proportional to $\theta^2$.
 The second term $W_\omega$ depends only on the two components of
 the gravitational fields $h^+$ and $h^-$ and is Weyl invariant.
 It is exactly the same as in the ref. [6].  The third term $W_\Omega$
is the new one. It stems from the term  linear in the momenta in the
 hamiltonian expression for the energy-momentum tensor. In the lagrangian formulation it is a
 total derivative and for manifolds without boundary, as we have
 supposed here, it vanishes. In that case, the only contribution of
 the hamiltonian $\theta$-term to the lagrangian is the
cosmological term. So, for manifolds without
boundary the complete effective action is
\be
W = -{\kappa_0c\over2}(W_L+W_\omega) \, .
\ee

\subsection{From the Virasoro to the trace anomaly}

 The central term in the Virasoro algebras changes the constraints
 from the FCC  to the SCC and breaks the diffeomorphism
 invariance. Using the extension of the general canonical method to the
 case when SCC are present [1], with the help of (1.5) and (3.12) we
 can easily find the variation of the  effective action
\bea
\delta W&=&-\int d\tau d\sigma\int d {\bar
 \sigma}[h^+(\sigma)\Delta_+(\sigma,{\bar\sigma})\varepsilon_-
 ({\bar\sigma})+h^-(\sigma)\Delta_-(\sigma,{\bar\sigma})\varepsilon_+({\bar\sigma})]\nonumber \\
 &=&\kappa_0c\int
 d^2\xi[\varepsilon_-(h^+)^{\prime\prime\prime}-\varepsilon_+(h^-)^{\prime\prime\prime}]\, .
\eea

 It is possible to add finite local counterterm, depending only on
 the gravitational fields $\Delta W(h^+,h^-, F)$, and to shift anomaly.
 Here we want to obtain the reparameterization invariance, so we
 require
\be
\delta\Delta W=-\delta W \, .
\ee

 It is easy to find expression for $\Delta W$ from our previous derivations. In
 (4.30) we just put $c\to -c, \theta \to 0$, and the most important $\phi \to
 2F$ in order to obtain the energy-momentum tensor for $\Delta W$.
 We change the sign of $c$ to ensure (4.41) and expunge $\theta$
 term for simplicity. There are two reasons for substituting
 auxiliary  field $\phi$ with the conformal part of the metric $2F$.
 First, the local counterterm $\Delta W$ must depend only on the metric components and
 second the field $\phi \over 2$ has  desirable  transformation properties.
From the first transformation (1.2), for $R\to {\phi \over 2}$, we obtain
\be
\delta {\phi \over 2}=-{1\over 2}(\varepsilon_++\varepsilon_-)^\prime
+(\varepsilon_+-\varepsilon_-){(h^-+h^+)^\prime\over
2(h^--h^+)}-{1\over
\sqrt{2}}(\varepsilon_-{\hat\partial}_+-\varepsilon_+{\hat\partial}_-){\phi \over 2} \,  ,
\ee
which for $\varepsilon^\pm =\varepsilon^1
-h^\pm\varepsilon^0$ and $\phi \to 2F$ gives
\be
\delta F =-\partial_1\varepsilon^1+{1\over
2}(h^-+h^+)\partial_1\varepsilon^0-\varepsilon^\mu\partial_\mu F \, .
\ee
This equation together with (2.14) completes the
transformation of the metric tensor (2.3) (see last ref. [3]), yielding the well known
expression
\be
\delta g^{\mu\nu}=g^{\mu\rho}\partial_\rho\varepsilon^\nu
+g^{\nu\rho}\partial_\rho\varepsilon^\mu
-\varepsilon^\rho\partial_\rho g^{\mu\nu} \, .
\ee

Then, in analogy with (4.34) we  obtain
\be
\Delta W = {\kappa_0c\over
 2}[W_L(2F,h^+,h^-)+W_\omega(h^+,h^-)] \, ,
\ee
and the complete effective action is
\be
W+\Delta W = - {\kappa_0c\over  2}W_L(g_{\mu\nu},\varphi) \, ,
\ee
with
\bea
W_L(g_{\mu\nu},\varphi)&=&W_L(\phi , h^+, h^-)-W_L(2F, h^+, h^-)\nonumber \\
 &=&\int d^2\xi\sqrt{-g}[\partial_+\varphi\partial_-\varphi+
 R\varphi+M^2(e^\varphi-1)] \, .
\eea
Here we have used (2.3) as well as its consequence $\sqrt{-g} R =\sqrt{-{\hat
g}} ({\hat R} -4 {\hat\nabla}_+ {\hat\partial}_- F)$
and introduced new field
\be
\varphi = \phi-2F \, .
\ee

From (4.42) and (4.43) we can find that $\varphi$ is a scalar field $\delta\varphi
=-\varepsilon^\mu\partial_\mu\varphi$, because the $\phi$ and  $F$ independent terms
are canceled. Note that from the same reason $W_\omega$ term disappeared.

 The new expression for the Liouville action (4.47) is manifestly
 reparameterization invariant, according to our construction. To achieve this,
 we  have introduced the field $F$, conformal part of the metric. Under Weyl
 rescaling $g_{\mu\nu}\to e^{2\sigma}g_{\mu\nu}$ we have $F\to
 F+\sigma$ while  $h^\pm$ are invariant. So, from (4.47) we can find
\be
\delta W_L = 2\sqrt{-g}(R+M^2)\sigma \,  ,
\ee
which is known expression for the trace anomaly.

 Now, we are ready to come back to the equation (3.1) and find the
 effective action for our model. Both parts, the scalar
 and the FP one, have the same form of the effective action (4.46)
 proportional to the Liouville action (4.47). Only the central
 charges are different. From (3.13) we have $c^x = D$ and $c^{gh} = -26$ so that
\be
W_{eff} = -{\hbar\over 48\pi}(D-26)W_L(g_{\mu\nu},
 \varphi) \, ,
\ee
where we put already obtained expression for $\kappa_0={\hbar\over 24\pi}$.

 The coefficient in front of the Liouville action is just ${1\over 2}$ of the
 central term for total energy-momentum tensor
\be
T_\pm = T_\pm^x+T_\pm^{gh} \, .
\ee
In the critical dimension $D = 26$, the constraints
$T_\pm$ become FCC  and the anomaly disappear. Originally, the expression
 for the induced gravity has been obtained
 from trace anomaly. Here, we calculated reparameterization
 anomaly, and only after shifting it with the counterterm $\Delta
 W$ we obtain the result of ref. [4].

 \section{Conclusion}

 In this paper we derived the Liouville theory as gravitational Wess-Zumino term
 using hamiltonian method.  This approach explicitly shows why the result, up
 to coefficient in front of the effective action, does not depend
 on the starting theory. It only depend on a sort of the symmetry,
 and here as consequence of the reparameterization invariance we got
 the Virasoro algebras of the constraints and Liouville action as
 an effective action. The trace of the original theory is in
 the central charge, which measures the numbers of the new degrees of
 freedom obtained  after quantization, and it appears as a coefficient in front of Liouville
 action in the effective action. Here we have two examples: the
 scalar and the ghost fields. They have different structure of the energy-momentum
 tensors: the square of the currents bilinear in the momenta for scalar
 fields (2.7) and linear in the momenta for the ghosts (2.23).

In Sec. II we fixed the reparameterization invariance in the
Polyakov string theory and made the canonical analysis for both
scalar and ghost part. Then we performed the quantization of the
matter and ghost fields in sec. III, obtaining the central charges
in the algebra of constraints, which change the nature of
constraints from the FCC to the SCC.

Sec. IV contains the main results of the paper. Here we proved that the
bosonic representation of Virasoro algebra, with arbitrary central
charge, must have a constraints of the form (4.30). They differ
from the standard expressions by a new term linear in the momenta.
We constructed the new bosonic theory which is equivalent to the
quantum theory of the original action. For the ghost part it is
bosonization, but for the matter field it is a non trivial
procedure because of  the presence of anomaly. The hamiltonian
$\theta$ term yielded two terms in the lagrangian formulation, one
was a total derivative and the other contributed to the mass term.
The first one could be important for the manifolds with boundary,
which we will investigate in the separate article. Note that even
in the case when the $\mu$-term is absent in the hamiltonian
approach the lagrangian will contain the mass term through the
$\theta$-part.

According to our normal ordering prescription, which has the role
of the regularization scheme we obtained the reparameterization
anomaly. In order to recover the diffeomorphism invariance we
added the finite local counterterm, which effectively changed the
regularization scheme and shifted anomaly to the trace one.

Our final results (4.50) is a complete and independent derivation
of the known Polyakov result, using hamiltonian method.

The canonical method, introduced in [1] for non-Abelian gauge
symmetry and developed here for reparameterization symmetry could be
applied to the supersymmetry and to the case where gauge fields
and metric are dynamical variables. We will publish it separately.


\begin{thebibliography}{99}
 \bibitem{1.}{ } B. Sazdovi\'c, Phys. Rev. {\bf D62} (2000) 045011.
 \bibitem{2.} { } M. Henneaux and S. Teitelboim, {\it Quantization of gauge
 systems} (Princeton Univ. Press, 1992).
 \bibitem{3.} { } A. Mikovi\'c and B. Sazdovi\'c, Mod. Phys. Lett. {\bf A10} (1995) 1041;
 {\bf A12} (1997) 501; M. Blagojevi\'c, D. S. Popovi\'c and B.
 Sazdovi\'c, Mod. Phys. Lett. {\bf A13} (1998) 911; Phys. Rev. {\bf
 D59} (1999) 044021.
 \bibitem{4.} { } A. M. Polyakov, Phys. Lett. {\bf B103} (1981) 207.
 \bibitem{5.} { } D. Friedan, Les Houches, Session XXXIX, 1982-{\it Recent
 advances in Field Theory and Statistical Mechanics}; O. Alvarez,
 Nucl. Phys. {\bf B216} (1983) 125.
 \bibitem{6.} { } T. Fujiwara, Y. Igarashi, J. Kubo and K. Maeda,
 Nucl. Phys. {\bf B391} (1993) 211; T. Fujiwara, T. Tabei, Y. Igarashi, K. Maeda and J.
 Kubo, Mod. Phys. Lett. {\bf A8} (1993) 2147; T. Fujiwara, Y. Igarashi, J. Kubo and T. Tabei,
 Phys. Lett. {\bf B336} (1994) 157.
 \bibitem{7.} { } C. Ford and L. O' Raifeartaigh, Nucl. Phys. {\bf B460} 203 (1996).

 \end{thebibliography}
\end{document}